\documentstyle[preprint,aps,epsf,floats,eqsecnum,tighten]{revtex}
\begin{document}

\preprint{\vbox{\hbox{JHU-TIPAC-97010}\hbox{UTPT--97-09}
		\hbox{hep-ph/9708327}\hbox{August, 1997}}}

\title{Hadronic Spectral Moments in Semileptonic $B$ Decays\\ With a Lepton Energy Cut}

\author{Adam F.~Falk$^a$ and Michael Luke$^b$}

\address{\medskip (a) Department of Physics and Astronomy, The Johns Hopkins
University\\ 3400 North
Charles Street, Baltimore, Maryland 21218 U.S.A.\\
{\tt falk@jhu.edu}\\ \medskip
(b) Department of Physics, University of Toronto\\ 60 St.~George
Street, Toronto, Ontario, Canada M5S 1A7\\
{\tt luke@medb.physics.utoronto.ca}}

\maketitle

\begin{abstract}%
We compute the first two moments $\langle (s_H-\overline{m}_D{}^2)^{1,2}\rangle$ of the final hadronic invariant mass in the inclusive decay $B\to X_c\,\ell\,\bar\nu$, in the presence of a cut $E_\ell^{\rm min}$ on the charged lepton energy.  These moments may be measured directly by experiments at the $\Upsilon(4S)$ using the neutrino reconstruction technique, which requires such a cut.  Measurement of these moments will place constraints on the nonperturbative parameters $\bar\Lambda$ and $\lambda_1$, which are relevant for extracting the quark masses $m_b$ and $m_c$, as well as the CKM angle $V_{cb}$.  We include terms of order $\alpha_s^2\beta_0$ and $1/m_b^3$ in the operator product expansion, and use the latter to estimate the theoretical uncertainty in the extraction of $\bar\Lambda$ and $\lambda_1$.
\end{abstract}


\newpage

\section{Introduction}

Semileptonic $B$ decays are a rich laboratory in which to study the bound state
structure of the $B$ meson.   Of particular importance are the nonperturbative parameters
$\bar\Lambda$, $\lambda_1$ and $\lambda_2$, which arise in the heavy quark expansion
for the meson mass $m_B$ to relative order $1/m_b^2$,
\begin{equation}
\label{massexpand}
  m_B=m_b+\bar\Lambda-{\lambda_1+3\lambda_2\over2m_b}
  +{\cal O}(\Lambda_{\rm QCD}^3/m_b^2)\,,
\end{equation}
where
\begin{eqnarray}
  \lambda_1&=&\langle B|\,\bar h_v(iD)^2 h_v\,|B\rangle/2m_B\,,\nonumber\\
  \lambda_2&=&\langle B|\,\bar h_v{g\over2}\sigma^{\mu\nu}G_{\mu\nu}
  h_v\,|B\rangle/6m_B\,.
\end{eqnarray}
Here $h_v$ is the heavy quark field, and $m_b$ is the $b$ quark pole mass.
Since these quantities also parameterize the inclusive semileptonic and radiative decay rates of the $B$ meson to
order
$1/m_b^2$\cite{CGG,BSUV,MW,Mannel,rare}, an accurate determination of them is essential for
a reliable extraction of  the CKM angle $|V_{cb}|$ from inclusive semileptonic $B$
decays.  The parameter
$\lambda_2$ is the expectation value of the leading operator which breaks heavy quark
spin symmetry and therefore may be determined from the $B$--$B^*$ mass difference,
yielding $\lambda_2\approx0.12\,{\rm GeV}^2$.  However, $\bar\Lambda$ and $\lambda_1$
cannot be determined solely from mass measurements.

There has been much recent interest in using inclusive observables other than the full semileptonic decay width to extract $\bar\Lambda$ and $\lambda_1$.  An analysis of the decay based on the heavy quark expansion yields an expression for the doubly differential decay rate $d\Gamma/ dq^2 d(v\cdot q)$, where $q^\mu$ is the total momentum of the leptons and $v^\mu$ is the four-velocity of the $B$ meson.  Any observable which may be constructed from this doubly differential rate is sensitive to a linear combination of $\bar\Lambda$ and $\lambda_1$.  Actually, the simplest differential distribution to measure experimentally is the energy spectrum of the charged  lepton, $ d\Gamma/ dE_\ell$, which is somewhat more complicated because it depends on more than just $q^\mu$.  Study of this distribution has already yielded useful constraints on $\bar\Lambda$ and $\lambda_1$ \cite{Vol,GKLW,GK,GS}.

In two previous publications~\cite{FLS}, we suggested that moments of the hadronic invariant  mass in semileptonic $B$ decay would also be interesting to study.  If $s_H$ is the invariant mass of the hadrons produced in the semileptonic  decay, and $\overline m_D=(m_D+3m_{D^*})/4$ is the spin-averaged $D$ meson mass, then positive moments of $s_H-\overline m_D^2$ vanish in the parton model at tree level.  Hence they are particularly sensitive to the power corrections proportional to $\bar\Lambda$ and $\lambda_1$.  However, it is difficult to measure these moments directly.  Until recently, only fairly weak bounds on $\langle (s_H-\overline m_D^2)^n\rangle$ could be obtained, by combining information on various exclusive decay channels~\cite{FLS}.  With the introduction of the technique of neutrino reconstruction, this situation is changing~\cite{CLEO96}.  Soon it will be possible to measure $s_H$ directly and inclusively, by reconstructing the neutrino momentum $p_\nu^\mu$ and using it to find $q^2$ and $v\cdot q$.  The only complication is that this reconstruction requires a number of constraints on the phase space of the leptons, most importantly a lower cut on the charged lepton energy of about 1.5~GeV~\cite{Thorndike}.  

While one might consider extrapolating the data to lower $E_\ell$ and measuring the moments of $s_H$ that way, it is clearly preferable to update the computation of the moments of $s_H$ to include this lepton energy cut {\it ab initio.}  It is the purpose of this note to present the results of such an analysis.  There are few new theoretical issues which arise in this case, although the actual calculation is considerably more complicated than without the cut.  We will refer the reader to our previous papers for a more complete exposition of the theory~\cite{FLS}, and only stress points where the new analysis differs in interesting ways from the old.  We will also extend our earlier analyses by including the complete power corrections to the moments to $O(1/m_b^3)$.  Dimensional estimates of the size of these corrections will help us to estimate the error in the extraction of $\bar\Lambda$ and $\lambda_1$ due to $1/m_b^3$ effects.  They will turn out to be under control for the first moment, but very large for the second moment, compromising its usefulness for obtaining constraints on $\bar\Lambda$ and $\lambda_1$.

\section{The Calculation}
\subsection{Power Corrections}

The analysis of the moments of $s_H$ begins with the doubly differential spectrum
$ d\Gamma/ d\hat q^2 d(v\cdot\hat q)$, where $\hat q^2=q^2/m_b^2$ and
$v\cdot\hat q=v\cdot q/m_b$.  This distribution may be written as a product of a
hadron tensor $T^{\mu\nu}$ and a lepton tensor $L^{\mu\nu}$,
\begin{equation}
   {d\Gamma\over d\hat q^2 d(v\cdot\hat q)}=
   T^{\mu\nu}(v,\hat q)\,L_{\mu\nu}(v,\hat q)\,.
\end{equation}
Each of these tensors has a decomposition in terms of scalar invariants, such as
\begin{eqnarray}
  L^{\mu\nu}&=&\left(-g^{\mu\nu}+{\hat q^\mu\hat q^\nu\over\hat q^2}\right)
  L_1+\left(v^\mu v^\nu+
  {\hat q^\mu\hat q^\nu(v\cdot\hat q)^2\over\hat q^4}
  -{v\cdot\hat q(v^\mu\hat q^\nu+v^\nu\hat q^\mu)\over\hat q^2}\right)
  L_2\nonumber\\
  &&\qquad- i\epsilon^{\mu\nu\alpha\beta}v_\alpha\hat q_\beta L_3\,.
\end{eqnarray}
The $L_i$ are scalar functions of $\hat q^2$ and $v\cdot\hat q$.  Here we neglect the masses of the leptons. If we impose a cut $E_\ell\ge xm_b$ on the charged lepton energy, then the effect is simply to modify the components $L_i$,
\begin{eqnarray}
  L_1&=&{\hat q^2\over24\pi[(v\cdot\hat q)^2-\hat q^2]^{3/2}}\bigg(
  4(v\cdot\hat q)^3-3\hat q^2v\cdot\hat q+4[(v\cdot\hat q)^2-\hat q^2]^{3/2}
  +6x\hat q^2\nonumber\\
  &&\qquad\qquad\qquad\qquad\qquad
  -12x(v\cdot\hat q)^2+12x^2v\cdot\hat q-8x^3\bigg)\,,\nonumber\\
  L_2&=&{\hat q^4\over8\pi[(v\cdot\hat q)^2-\hat q^2]^{5/2}}
  (2x-v\cdot\hat q)(\hat q^2-4xv\cdot\hat q+4x^2)\,,\nonumber\\
  L_3&=&{\hat q^2\over8\pi[(v\cdot\hat q)^2-\hat q^2]^{3/2}}
  (\hat q^2-4xv\cdot\hat q+4x^2)\,.
\end{eqnarray}
The cut may be removed by setting it equal to the minimum
charged lepton energy, $x\to{1\over2}[(v\cdot\hat q)^2-\hat q^2]^{1/2}$, in
which case $L_1=\hat q^2/3\pi$, $L_2=L_3=0$.  Expressions for the analogous
hadron tensor components $T_i$ may be found in Refs.~\cite{BSUV,MW,GK}.  The hadron tensor is independent of the cut $x$.

Since the hadron tensor $T^{\mu\nu}$ is computed with operator product
expansion techniques which assume parton-hadron duality, the calculation must be
smeared by integration over at least one of $\hat q^2$ or $v\cdot\hat q$ before
meaningful observables may be extracted~\cite{BSUV,MW}.  A nonzero cut $x$ has the effect
of restricting the available phase space for the leptons, which controls the range
of integration in $\hat q^2$ and $v\cdot\hat q$.  If $x$ is too large, the
integration is too restricted, and the operator product expansion breaks down.This is known to happen, for example, near the lepton endpoint in charmless
semileptonic $B$ decays, for $E_\ell$ above about 2.2~GeV~\cite{BSUV,MW,NB,BM}. However, our calculations of the coefficients of the $1/m_b^3$ corrections indicate
that the lepton energy cut of 1.5~GeV which is required for the neutrino reconstruction
technique is not severe enough to cause such problems.

The next step is to compute parton level moments of the form $\langle\hat E_0^m(\hat s_0-\hat
m_c^2)^n\rangle$, where $\hat s_0$ and $\hat E_0$
are the invariant mass and total energy of the strongly interacting partons
produced in the semileptonic decay of the $b$ quark.  The energy is computed in
the $b$ rest frame.  The partonic variables are related implicitly to the
hadronic variable $s_H$ by
\begin{eqnarray}
  \hat E_0&=&1-v\cdot\hat q\,,\qquad\hat s_0=1-2v\cdot\hat q+\hat q^2\,;\nonumber\\
  s_H&=&m_B^2-2m_Bv\cdot q+q^2\,.
\end{eqnarray}
Since these expressions involve both $m_B$ and $m_b$, they must be inverted order
by order in the heavy quark expansion using Eq.~(\ref{massexpand}).  The final result will be moments of the
form $\langle(s_H-\overline m_D^2)^m\rangle$, for $m=1,2$.  In fact, we will see that
only the first moment is really reliable, where the terms of order $\Lambda_{\rm
QCD}$ and $\Lambda^2_{\rm QCD}$ in the heavy quark expansion are known.  By contrast,
the second moment starts only at order $\Lambda^2_{\rm QCD}$, and hence is extremely
sensitive to the large number of unknown parameters which arise at order
$\Lambda^3_{\rm QCD}$.

\subsection{Radiative Corrections}

While it is possible to calculate the radiative corrections by calculating the
$O(\alpha_s)$ contributions to the $T_i$'s themselves, it is much simpler to calculate directly the leading corrections to the parton model rate.
The only subtlety in the calculation arises in determining the boundaries of phase
space when the electron cut is imposed. Since the limits of integration of the electron
energy depend on the parton level invariant mass $s_0$ and lepton invariant mass
squared $q^2$, for given values of $s_0$ and $q^2$ the electron cut may either lie below
the lower limit of integration, in the region of integration, or above the upper limit
of integration.  Let $f(s_0,q^2)$ be any smooth weighting function.  Then the phase space integral is divided into 
kinematic regions, depending on the values of $x$, $s_0$ and $q^2$,
\begin{itemize}
\item For $x<{1\over 2}(1-m_c)$,
\begin{eqnarray}
  \langle f(s_0,q^2)\rangle&=&\int_{m_c^2}^{(1-2x)^2}\, ds_0\,
  \left[\int_0^{2x(1-s_0-2x)\over 1-2x}\,dq^2\,f(s_0,q^2){d \Gamma_1\over d q^2 ds_0}\right.\nonumber \\ &&\qquad\qquad\qquad\qquad+\left.
  \int_{2x(1-s_0-2x)\over 1-2x}^{(1-\sqrt{s_0})^2}\,dq^2\,f(s_0,q^2){d \Gamma_2\over d q^2
  ds_0}\right]\nonumber\\&&+\int_{(1-2x)^2}^{1-2x}\,
  ds_0\,\int_0^{2x(1-s_0-2x)\over 1-2x}\,dq^2\,f(s_0,q^2){d \Gamma_1\over d q^2
  ds_0}\,,\nonumber
\end{eqnarray}
\item For $x>{1\over 2}(1-m_c)$,
\begin{eqnarray}
  \langle f(s_0,q^2)\rangle &=&\int_{m_c^2}^{1-2x}\,
  ds_0\,\int_0^{2x(1-s_0-2x)\over 1-2x}\,dq^2\,f(s_0,q^2){d
  \Gamma_1\over d q^2 ds_0}\,.\nonumber
\end{eqnarray}
\end{itemize}
Here $d \Gamma_1/d q^2 ds_0$ is the differential rate calculated with the electron
energy cut imposed, and $d \Gamma_2/d q^2 ds_0$ is the differential rate calculated with no cut, corresponding to $x$ lying below the lower limit of integration for
the electron energy.  In the regions of phase space omitted from the expressions
above, the cut lies above the upper limit of integration for the electron energy.

The ``BLM -enhanced"~\cite{BLM} two-loop corrections are those which are proportional to $\alpha_s^2\beta_0$, where $\beta_0=11-2n_f/3$ is the first term in the QCD beta function.  These corrections dominate the two-loop corrections to many
processes in QCD, and their effects in $b$ decays have been discussed extensively in the literature\cite{blmpapers}.  They are straightforward to calculate numerically using the techniques of Ref.~\cite{smvol}, and no new subtleties are introduced into the calculation when an electron energy cut is added.  Because of the renormalon ambiguity in its definition \cite{renorms}, $\bar\Lambda$ is only defined order by order in perturbation theory.  Since we are including the $\alpha_s^2\beta_0$ terms in our extraction of $\bar\Lambda$, the resulting value is the ``two-loop" $\bar\Lambda$, and should only be compared with other extractions of $\bar\Lambda$ at the same order. 

\section{Results and Discussion}

The expansion of a moment of $s_H$ takes the following general form, up to
terms of relative order $1/m_b^3$:
\begin{eqnarray}
  \langle(s_H-\overline m_D^2)^m\rangle&=&m_B^{2m}\Bigg\{
  C^{(m)}_{1}{\alpha_s(m_b)\over\pi}+\left[C^{(m)}_{22}\beta_0+C^{(m)}_{21}\right]
  {\alpha^2_s(m_b)\over\pi^2}
  +{\cal O}(\alpha_s^3(m_b))\nonumber\\
  &&+D^{(m)}_1{\bar\Lambda\over\overline m_B}+
  \left[D^{(m)}_{20}{\bar\Lambda^2\over\overline m_B^2}
  +D^{(m)}_{21}{\lambda_1\over\overline m_B^2}
  +D^{(m)}_{22}{\lambda_2\over\overline m_B^2}\right]\\
  &&+\left[D^{(m)}_{30}{\bar\Lambda^3\over\overline m_B^3}
  +D^{(m)}_{31}{\bar\Lambda\lambda_1\over\overline m_B^3}
  +D^{(m)}_{32}{\bar\Lambda\lambda_2\over\overline m_B^3}
  +D^{(m)}_{33}{\rho_1\over\overline m_B^3}
  +D^{(m)}_{34}{\rho_2\over\overline m_B^3}\right.\nonumber\\
  &&\left.+\sum_{i=1}^4 T^{(m)}_{3i}{{\cal T}_i\over\overline m_B^3} \right]\Bigg\}
  \nonumber\,.
\end{eqnarray}
All the coefficients which appear are functions of the lepton energy cut
$E_\ell^{\rm min}$.  The parameters $\rho_1$ and $\rho_2$ are expectation values of
local operators of dimension six which arise at order $1/m_b^3$ in the heavy quark
expansion, 
\begin{eqnarray}
  & \langle B|\bar{h}_v(iD_\alpha)(iD_\mu)(iD_\beta)h_v|B\rangle
  \equiv{1\over 3} \rho_1\left(g_{\alpha\beta} - v_\alpha v_\beta \right) v_\mu\,,
  &   \nonumber\\
  & \langle B^{(*)}|\bar{h}_v(iD_\alpha)(iD_\mu)(iD_\beta)\,\gamma_\delta\gamma_5
  \,h_v|B^{(*)}\rangle \equiv{1\over 6}d_H\rho_2 
  \,i\epsilon_{\nu\alpha\beta\delta}v^\nu
  v_\mu\,,&
\end{eqnarray}
where $d_H=3$ and $d_H=-1$, respectively, for matrix elements between $B$ and  $B^*$ states.  The ${\cal T}_i$ are related to nonlocal time ordered products of $1/m_b^2$ terms in the operator product expansion with $1/m_b$ terms in the Lagrangian~\cite{BM},
\begin{eqnarray}\label{Tproducts}
  \langle B^{(*)}|\bar h_v\, (i D_\bot)^2\, h_v\, i\int d^3x\int_{-\infty}^0 dt\,
  {\cal L}_{I}(x)|B^{(*)}\rangle+{\rm h.c.}&\equiv&\frac{{\cal T}_1+d_H
  {\cal T}_2}{m_b}, \\ \nonumber
  \langle B^{(*)}|\bar h_v\, {g\over2}\,\sigma_{\mu\nu}\, G^{\mu\nu}\,
  h_v\, i\int d^3x\int_{-\infty}^0 dt\, {\cal L}_{I}(x)|B^{(*)}\rangle+
  {\rm h.c.}&\equiv&
  \frac{{\cal T}_3+d_H{\cal T}_4}{m_b}.
\end{eqnarray}
The parameters $\rho_i$ and ${\cal T}_i$ are determined by nonperturbative QCD, and their values are not known; however, we compute their coefficients to ensure that
none of them are anomalously large.   We will use dimensional analysis to
estimate $\rho_i$ and ${\cal T}_i$, to obtain a rough estimate of the error in the extraction
of $\bar\Lambda$ and $\lambda_1$ induced by $1/m_b^3$ effects. 

The coefficients $D_{ij}$ are
themselves functions of $\alpha_s$, although we will not compute radiative corrections
to any of these coefficients.  At present,
$D_1^{(m)}$ are known to order $\alpha_s$ only in the absence of an electron energy cut.  While
it would certainly be desirable to include terms of order $\alpha_s\bar\Lambda/m_b$ for general $E_\ell^{\rm min}$, the calculation is
quite difficult and we have not attempted it here.  
This omission is particularly
important for the second moment, since $D^{(2)}_1=0$.  Without the cut, the term in question
is numerically as large as the leading terms proportional to
$C^{(2)}_1$ and $D^{(2)}_{20}$~\cite{GK}.\footnote{The authors of Ref.~\cite{GK}
correct a numerical error in this term in Ref.~\cite{FLS}.}  (In the case of the first
moment, by contrast, the radiative correction to $D^{(1)}_1$ is only a few percent~\cite{FLS}.)  For
this reason, as well as because of the large $1/m_b^3$ corrections,  it is dangerous to use the second moment in a measurement of
$\bar\Lambda$ and $\lambda_1$.  We will include our ignorance of these terms in the 
estimate of the theoretical error in our results below.

\begin{figure}[tbh]
\epsfxsize=12 cm
\hfil\epsfbox{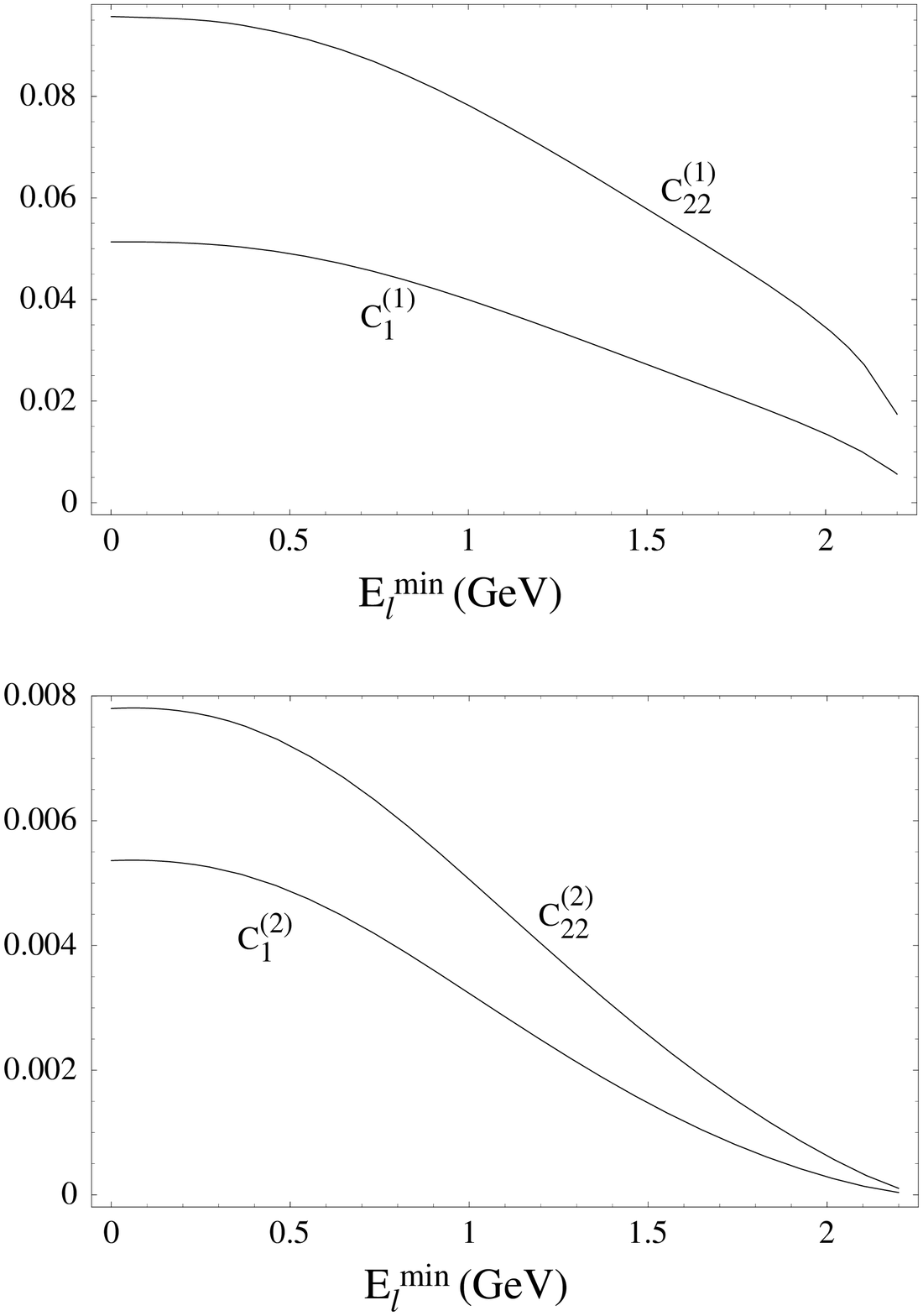}\hfill
\vskip0.4cm
\caption{Coefficients $C_{ij}^{(k)}$ of the radiative corrections to the first and
second moments, as functions of the electron energy cutoff in GeV.}
\label{cplots}
\end{figure}

\begin{figure}[tbh]
\epsfxsize=12 cm
\hfil\epsfbox{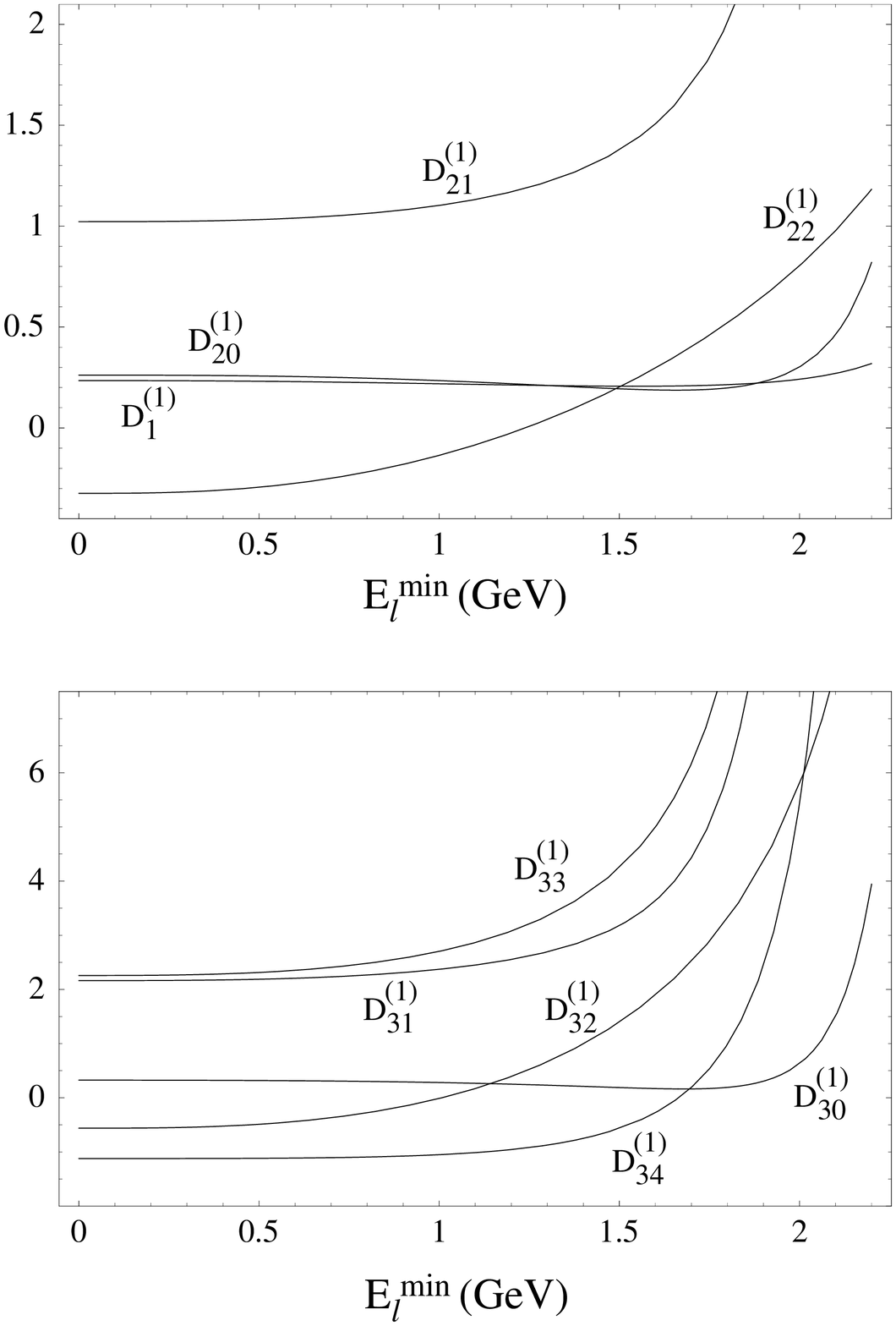}\hfill
\vskip0.4cm
\caption{Coefficients $D_{ij}^{(1)}$ of the power corrections to the first moment of
the hadronic invariant mass spectrum, as functions of the electron energy cutoff in
GeV.}
\label{d1plots}
\end{figure}

\begin{figure}[tbh]
\epsfxsize=12 cm
\hfil\epsfbox{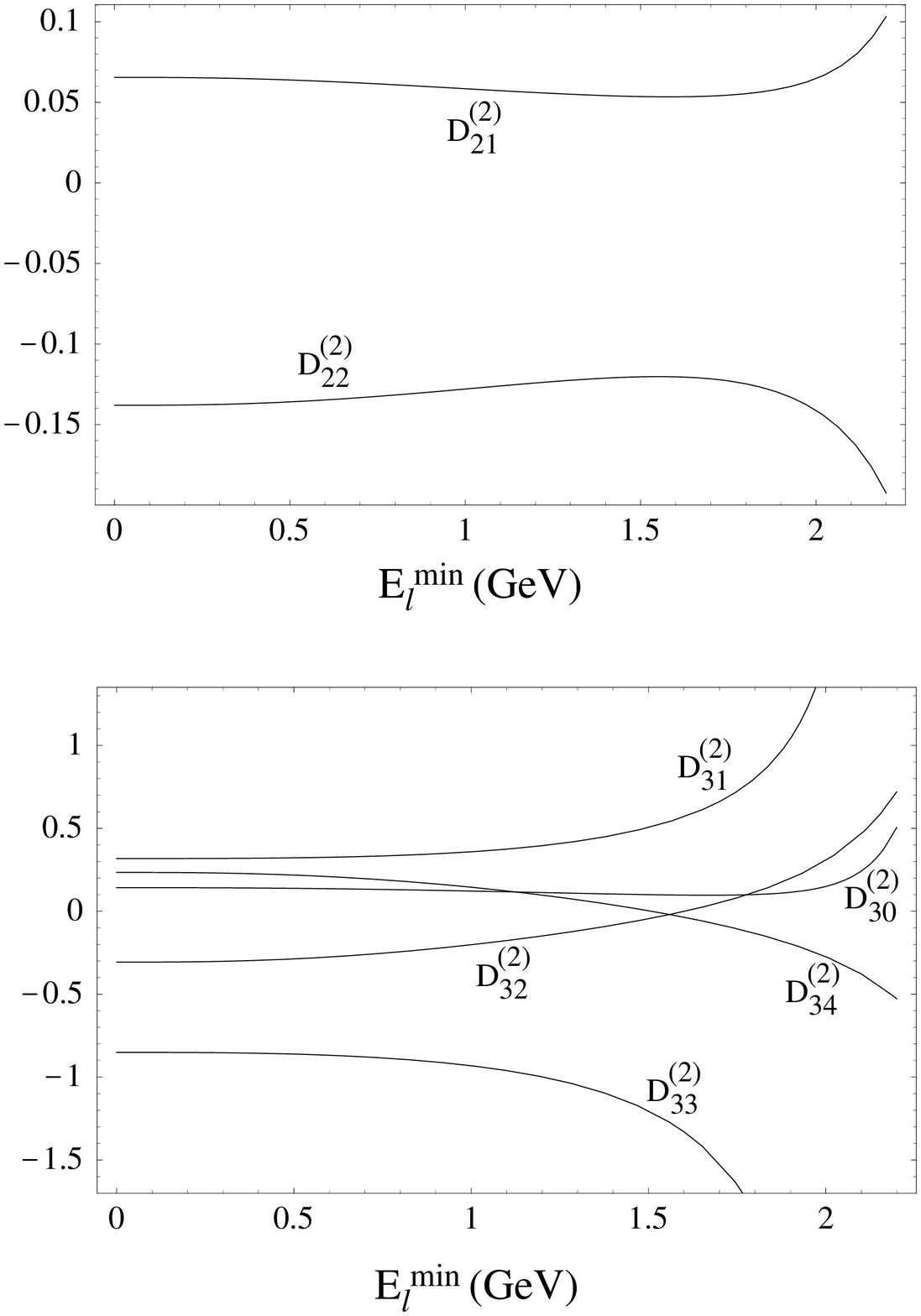}\hfill
\vskip0.4cm
\caption{Coefficients $D_{ij}^{(2)}$ of the power corrections to the second moment of
the hadronic invariant mass spectrum, as functions of the electron energy cutoff in
GeV.}
\label{d2plots}
\end{figure}

Let $y=E_\ell^{\rm min}/m_B$ be the scaled lepton energy cut.  When $y=0$, we
reproduce the known results~\cite{GK,FLS},
\begin{equation}
\begin{array}{l@{\extracolsep{.5cm}}l@{\extracolsep{.5cm}}l@{\extracolsep{.5cm}}l
@{\extracolsep{.5cm}}l}
  C^{(1)}_1=0.051& C^{(1)}_{22}=0.096& & & \\
  D^{(1)}_1=0.23& D^{(1)}_{20}=0.26& D^{(1)}_{21}=1.0&  D^{(1)}_{22}=-0.32& \\
  D^{(1)}_{30}=0.33& D^{(1)}_{31}=2.2& D^{(1)}_{32}=-0.56&
  D^{(1)}_{33}=2.3& D^{(1)}_{34}=-1.2\\
  T^{(1)}_{31}=1.6& T^{(1)}_{32}=0.80&
  T^{(1)}_{33}=1.5& T^{(1)}_{34}=0.41&\\
\end{array}
\end{equation}
and
\begin{equation}
\begin{array}{l@{\extracolsep{.5cm}}l@{\extracolsep{.5cm}}l@{\extracolsep{.5cm}}l
@{\extracolsep{.5cm}}l}
  C^{(2)}_1=0.0054 & C^{(2)}_{22}=0.0078& & &\\
  D^{(2)}_1=0& D^{(2)}_{20}=0.066& D^{(2)}_{21}=-0.14&D^{(2)}_{22}=0& \\
  D^{(2)}_{30}=0.14& D^{(2)}_{31}=0.32& D^{(2)}_{32}=-0.31&  D^{(2)}_{33}=-0.85&
  D^{(2)}_{34}=0.23\\
  T^{(1)}_{31}=-0.14& T^{(1)}_{32}=-0.41&
  T^{(1)}_{33}=0& T^{(1)}_{34}=0\,.&\\
\end{array}
\end{equation}
The leading radiative correction to $D_1^{(1)}$ is $0.099\,\alpha_s(m_b)/\pi$, and to $D_1^{(2)}$ is $0.038\,\alpha_s(m_b)/\pi$.  For the preferred experimental cut of 1.5~GeV, for which $y=0.28$, we find
\begin{equation}
\begin{array}{l@{\extracolsep{.5cm}}l@{\extracolsep{.5cm}}l@{\extracolsep{.5cm}}l
@{\extracolsep{.5cm}}l}
  C^{(1)}_1=0.028& C^{(1)}_{22}=0.058& & &\\
  D^{(1)}_1=0.21&D^{(1)}_{20}=0.19& D^{(1)}_{21}=1.4&D^{(1)}_{22}=0.19&\\
  D^{(1)}_{30}=0.19& D^{(1)}_{31}=3.2& D^{(1)}_{32}=1.4&
  D^{(1)}_{33}=4.3& D^{(1)}_{34}=-0.56\\
  T^{(1)}_{31}=2.0& T^{(1)}_{32}=1.8&
  T^{(1)}_{33}=1.7& T^{(1)}_{34}=0.91&
\end{array}
\end{equation}
and
\begin{equation}
\begin{array}{l@{\extracolsep{.5cm}}l@{\extracolsep{.5cm}}l@{\extracolsep{.5cm}}l
@{\extracolsep{.5cm}}l}
  C^{(2)}_1=0.0015& C^{(2)}_{22}=0.0026& & &\\
  D^{(2)}_1=0& D^{(2)}_{20}=0.054&D^{(2)}_{21}=-0.12&  D^{(2)}_{22}=0&\\
  D^{(2)}_{30}=0.10& D^{(2)}_{31}=0.51& D^{(2)}_{32}=-0.045&
  D^{(2)}_{33}=-1.2& D^{(2)}_{34}=0.0032\\
  T^{(2)}_{31}=-0.12&T^{(2)}_{32}=-0.36&
  T^{(2)}_{33}=0& T^{(2)}_{34}=0\,.&
\end{array}
\end{equation}
We present plots of the coefficients $C_i$ and $D_i$ for arbitrary $E_\ell^{\rm min}$ in Figs.~\ref{cplots}--\ref{d2plots}.  Note that the power corrections tend to blow up as the electron
cutoff approaches its maximum value;  fortunately, for a cut of 1.5 GeV the
coefficients are not dramatically larger than without a cut.

\begin{figure}[tbh]
\epsfxsize=12 cm
\hfil\epsfbox{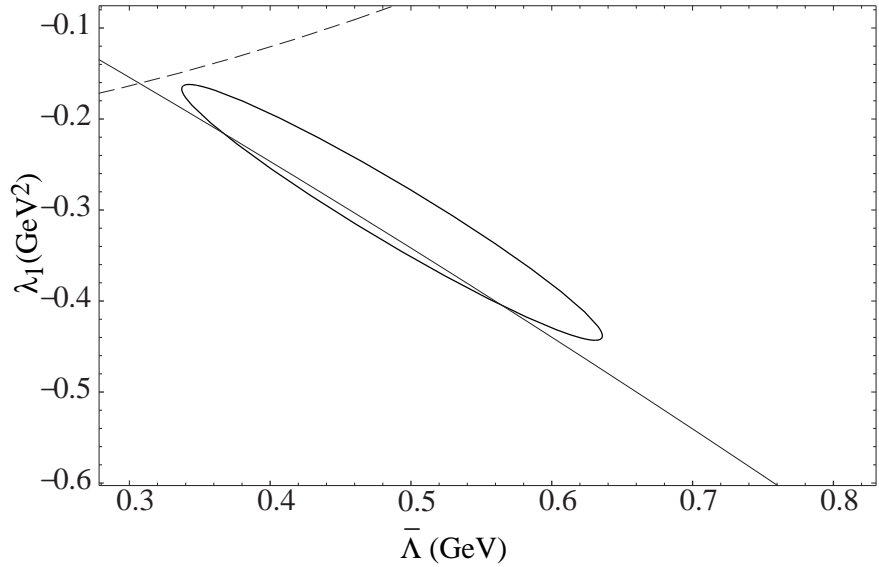}\hfill
\vskip0.4cm
\caption{Estimate of the theoretical uncertainty in $\bar\Lambda$ and $\lambda_1$ due
to unknown $1/m_b^3$ and $\alpha_s\bar\Lambda/m_b$ contributions to the moments (see text for description).  For the purpose of illustration, perfect experimental measurements of  $\langle s_H-\overline m_D^2\rangle=0.30\,{\rm GeV}^2$ 
and $\langle (s_H-\overline m_D^2)^2\rangle=0.96\,{\rm GeV}^4$ have been assumed.
The solid and dashed lines show the constraints on $\bar\Lambda$ and $\lambda_1$ from
the first and second moments, respectively, while the area in the shaded ellipse
shows the estimated allowed range.}
\label{errorplot1}
\end{figure}

The effects of the $1/m_b^3$ corrections to the moments in the extraction
of $\bar\Lambda$ and $\lambda_1$ are displayed in Fig.~\ref{errorplot1}.  For the
purpose of illustration, we first assume perfect
experimental measurements of $\langle s_H-\overline m_D^2\rangle=0.30\,{\rm GeV}^2$ 
and $\langle (s_H-\overline m_D^2)^2\rangle=0.96\,{\rm GeV}^4$, and then extract $\bar\Lambda$ and $\lambda_1$ using the theoretical expressions at order $1/m_b^2$.  We take $\alpha_s(m_b)=0.22$.  These values for the first and second moments restrict $\bar\Lambda$ and $\lambda_1$ to lie on the solid and dashed curves, respectively, meeting at the point $(\bar\Lambda,\lambda_1)=(0.31\,{\rm GeV}, -0.16\,{\rm GeV}^2)$.  The hypothesized data have been chosen so that this intersection coincides with the central values obtained from an existing analysis of the lepton energy spectrum~\cite{GKLW}.  We then may estimate the theoretical uncertainty in our hypothetical result by following the approach of Ref.~\cite{GK}.
By dimensional analysis, the parameters $\rho_i$ and ${\cal T}_i$ are
all of order $\Lambda_{\rm QCD}^3$, and an estimate of their effect on
the extraction of $\bar\Lambda$ and $\lambda_1$ is obtained
by varying their magnitudes independently in the range
$0-(0.5\,{\rm GeV})^3$.  Since the vacuum saturation approximation suggests
that $\rho_1>0$, we take it to be positive, and we eliminate $\rho_2$ by making use of the relation
between $\rho_2$, ${\cal T}_2$ and ${\cal T}_4$ and the $D^*$--$D$ and $B^*$--$B$ mass
splittings presented in Ref.~\cite{GK}.  
Finally, we vary the unknown coefficient of the $\alpha_s\bar\Lambda/m_b$ term for both moments
between half and twice its value with the cut removed.  
Varying the unknown parameters randomly in the allowed ranges, we find that the shaded ellipse shown in the figure and centered about the mean of the distribution contains 68\% of the points. (Because the $\bar\Lambda^3$, $\bar\Lambda\lambda_1$ and $\bar\Lambda\lambda_2$ terms bias the determination of $\bar\Lambda$ and $\lambda_1$ in a known way, the distribution is not centered about the point  extracted from the theory at order $1/m_b^2$.).
The region inside the ellipse gives a reasonable estimate of the theoretical 
error in the extraction of
$\bar\Lambda$ and $\lambda_1$ due to higher order effects.  Note that, as expected, the
constraints from the second moment are strongly affected by higher order corrections,
whereas the constraints from the first moment are quite tightly distributed about the
leading result.  Thus, as discussed earlier, only the linear combination
of $\bar\Lambda$ and $\lambda_1$ given by the first moment is 
significantly constrained.  

\begin{figure}[tbh]
\epsfxsize=12 cm
\hfil\epsfbox{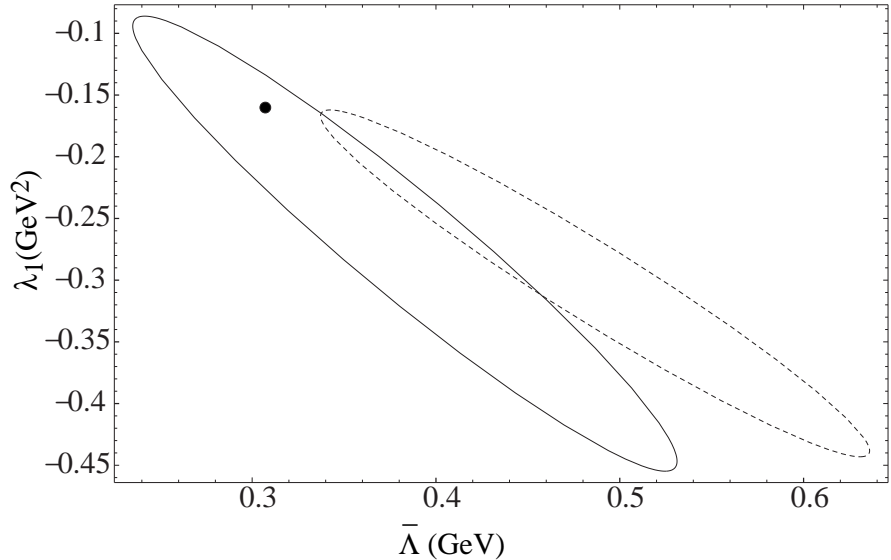}\hfill
\vskip0.4cm
\caption{Estimate of the theoretical uncertainty in $\bar\Lambda$ and $\lambda_1$ due
to unknown $1/m_b^3$ contributions from the shape of the
electron spectrum, using the
same approach as in the previous plot.  The central value at $O(1/m_b^2)$ is given by
the black dot, while the dashed ellipse is the plot from the previous
figure, shown here for comparison.  Only the relative sizes and orientations of the two ellipses are meaningful in this figure, not their relative positions.}
\label{errorplot2}
\end{figure}

Constraints on $\bar\Lambda$ and $\lambda_1$ also have been obtained from moments of the
lepton energy spectrum above 1.5~GeV~\cite{GKLW}.  To compare the theoretical errors in this approach to ours, we have performed an analysis analogous to that of the previous paragraph.  This is similar to what was done in Ref.~\cite{GK}, but we also include the terms proportional to $\alpha_s^2\beta_0$~\cite{GS}.  The result is shown in Fig.~\ref{errorplot2}.
The size of the ellipse from the lepton energy analysis
is slightly larger than, but comparable to, that which we obtained from the hadronic mass moments.  Of course, the relative position of the ellipses is meaningless, since the hadronic mass moments have not yet been measured.  Unfortunately, the two experiments effectively constrain the same linear combination of $\bar\Lambda$ and $\lambda_1$, so that the measurements cannot be combined to determine both parameters simultaneously.  Instead, an observable sensitive to a different linear combination
of $\bar\Lambda$ and $\lambda_1$, such as the first moment in the photon spectrum
in $B\rightarrow X_s\gamma$~\cite{KL}, will be required.  In the meantime, consistency of the various
allowed regions in the $\bar\Lambda$-$\lambda_1$ plane will provide a powerful test of the heavy quark expansion for these decays.

\section{Conclusions}

We have presented a calculation of the first two hadronic invariant mass moments in semileptonic $B$ decay, in the presence of a moderate cut on the energy of the charged lepton.  We included effects up to order $\alpha_s^2\beta_0$ and $1/m_b^3$.  These moments may be used to measure a linear combination of the HQET parameters $\bar\Lambda$ and $\lambda_1$, with a theoretical accuracy which is comparable to, or slightly better than, the accuracy obtained from an analysis of the charged lepton energy spectrum.  The consistency of the results obtained from these approaches will provide a test of the heavy quark expansion as applied to semileptonic $B$ decays.  To extract $\bar\Lambda$ and $\lambda_1$ simultaneously, it will be necessary to combine this analysis with that of a quantity sensitive to a different linear combination of the two parameters.

\acknowledgements
We are grateful to Martin Savage for his invaluable collaboration on closely related
work.  We also thank Mark Wise for useful discussions, and Ed Thorndike for alerting us
to the importance of such an investigation.  A.F.F.~was supported in part by the
National Science  Foundation under Grant No.~PHY-9404057 and National Young
Investigator Award No.~PHY-9457916, by the Department of Energy under Outstanding
Junior Investigator Award No.~DE-FG02-94ER40869, by the Alfred P.~Sloan
Foundation, and by the Research Corporation as a Cottrell Scholar.  M.L.~was supported by the Alfred P. Sloan Foundation and by the Natural
Sciences and Engineering Research Council of Canada.


\end{document}